\documentclass[epj]{svjour}
\usepackage{graphics}
\begin{document}
\title{Hawking radiation of Kerr-Newman-de Sitter black
hole}
\author{Li huaifan\inst{1,2}  Zhang Shengli\inst{1} Wu Yueqin\inst{2}, Zhang Lichun\inst{2} \and Zhao Ren\inst{2}% etc
\thanks{\emph{Present address:} Institute of Theoretical Physics, Shanxi
Datong University, Datong 037009 P.R.China}%
}                     % Do not remove
\institute{Department of Applied Physics, Xi'an Jiaotong University,
Xi'an 710049 P.R.China \and Institute of Theoretical Physics, Shanxi
Datong University, Datong 037009 P.R.China}
\date{Received: date / Revised version: date}
\abstract{We extend the classical Damour-Ruffini method and discuss
Hawking radiation in Kerr-Newman-de Sitter(KNdS) black hole. Under
the condition that the total energy, angular momentum and charge of
spacetime are conserved, taking the reaction of the radiation of the
particle to the spacetime and the relation between the black hole
event horizon and the cosmological horizon into consideration, we
derive the black hole radiation spectrum. The radiation spectrum is
no longer a pure thermal one. It is related to the change of the
Bekenstein-Hawking entropy corresponding the black hole event
horizon and the cosmological horizon. It is consistent with an
underlying unitary theory.
 % end of PACS codes
} %end of abstract
\maketitle
\section{Introduction}
\label{intro} In the 1970s, Hawking [1] pointed that the black hole
can radiate particles. Vacuum fluctuations near the surface of the
black hole would produce virtual particle pairs. When the virtual
particles with negative energy go into black hole via the tunnel
effect, the energy of the black hole will decrease. At the same
time, the particle with positive energy may thread out the
gravitation region outside the black hole. Equivalently, the black
hole radiates a particle. Gibbons and Hawking also demonstrated that
the energy spectrum of radiation is exactly thermal [2]. The Hawking
radiation demonstrates that the black hole is no longer an ultimate
state of a star and will evolve. Since Hawking did not consider the
reaction of the radiation to spacetime, the Hawking radiation
spectrum is an exact black body spectrum. So we can only obtain a
parameter-temperature from black body spectrum. Thus the black hole
radiation will not bring any information about matter in the black
hole. It means that if the black hole completely evaporates, all
black hole information including the unitary property will
disappear. The information loss of the black hole means that the
pure quantum state will decay to a mixed state. This violates the
unitary principle in quantum mechanics. This is a serious challenge
to the theoretical basic of quantum mechanics.

Before 2004, Hawking firmly believed that information is not
conservative during the black hole evolution. The black hole
evolution did not satisfy the unitary principle in quantum
theory[3]. However, some physicists advocate that information is
conserved during the black hole evolution and the unitary principle
should be satisfied [4]. However, the aforesaid two view points have
not been proved for 30 years. In 2004, during the 17th International
Conference on General Relativity and Gravitation, Hawking brought
about a tremendous conservation. He proposed that the information
should be conservative during the black hole formation and
evaporation process [5]. But Hawking did not proved strictly his
proposal.

In 2000, Parikh and Wilczek proposed a semiclassical method for
calculating the modified spectrum of the black hole Hawking
radiation [6]. In this method, the black hole Hawking radiation is
understood as a sort of quantum tunneling. The potential barrier is
determined by the energy of emission particles. The key of this
method is emphasizing energy conservation during the particle
emission process and establishing a good coordinate system at
horizon. Using this method Parikh and Wilczek have calculated the
emission modified spectrum of particles through a Schwarzschild
black hole and a Reissner-Nordstr\"om one. The result departs from
the purely thermal spectrum. It satisfies the unitary principle and
support the result of information conservation. Subsequently, the
Hawking radiation modified spectrums of axially symmetric black
holes have been calculated [7-35]. They all satisfy the unitary
theory and support the result of information conservation.

The core idea of Parikh and Wilczek's work is considering the total
energy conservation of spacetime in the process of black hole
radiation, and the energy of the black hole can produce
fluctuations. When Ref.[21-23] calculated the radiation spectrum of
axisymmetric black hole using the tunneling method proposed by
Parikh and Wilczek, they only considered that the energy and charge
of the black hole produce fluctuations. The change of the black hole
angular momentum is determined by the change of the black hole
energy. They did not consider the effect of the rotation of the
black hole radiated particles on the black hole angular momentum.

It is well-known that as a thermodynamic system the black hole has
temperature and entropy. For a spacetime that does not include a
cosmological term, the state parameters of this thermodynamic system
are all embodied on the black hole horizon surface. The black hole
event horizon is the "window" that transports information to outside
world. The radiation spectrum of this type black hole has been
discussed by the tunnel method. However, for the spacetime that
include a cosmological term, the state parameters of this
thermodynamic system embody on not only the black hole event horizon
surface but also the cosmological horizon surface. Both the black
hole event horizon and the cosmological horizon are the "windows"
that transport information to outside world. Because the black hole
event horizon and the cosmological horizon have the same state
parameters, there is a correlation between the black hole event
horizon and the cosmological horizon. At present when the two
horizons have relevance, the study on black hole radiation spectrum
has not yet reported.

In this paper, we extend the classical Damour-Ruffini method [36] to
discuss the radiation spectrum in the Kerr-Newman-de Sitter(KNdS)
spacetime. Under the condition that the spacetime total energy,
total charge and total angular momentum are conserved, we derive the
black hole radiation spectrums after considering the reaction of
radiation to the spacetime and the relation between the black hole
event horizon and the cosmological horizon. The radiation spectrum
is no longer a pure thermal spectrum and deviate from the black body
spectrum. It is related to the change of the Bekenstein-Hawking
entropy corresponding the black hole event horizon and the
cosmological horizon.
\section{Kerr-Newman-de Sitter black hole}
The Kerr-Newman-de Sitter(KNdS) spacetime line element [37,38]:

\[
ds^2 = - \frac{1}{\rho ^2}\left( {\Delta _r - \Delta _\theta a^2\sin
^2\theta } \right)dt^2 + \frac{\rho ^2}{\Delta _r }dr^2 + \frac{\rho
^2}{\Delta _\theta }d\theta ^2
\]

\[
+ \frac{1}{\rho ^2\Xi ^2}\left[ {\Delta _\theta (r^2 + a^2) - \Delta
_r a^2\sin ^2\theta } \right]\sin ^2\theta d\varphi ^2
\]

\begin{equation}
\label{eq1}
 - \frac{2a}{\rho ^2\Xi }\left[ {\Delta _\theta (r^2 + a^2) - \Delta _r }
\right]\sin ^2\theta dtd\varphi ,
\end{equation}

\noindent where

\[
\rho ^2 = r^2 + a^2\cos ^2\theta , \quad \Delta _\theta = 1 +
\frac{1}{3}\Lambda a^2\cos ^2\theta ,
\]

\begin{equation}
\label{eq2} \Delta _r = (r^2 + a^2)\left( {1 - \frac{1}{3}\Lambda
r^2} \right) - 2Mr + q^2, \quad \Xi = 1 + \frac{1}{3}\Lambda a^2.
\end{equation}
Here the parameters $M$, $a$ and $q$ are the associated with the
mass, angular momentum, and charge parameters of the space-time,
respectively, and $\Lambda $ is the positive cosmological constant.
Based on the metric (\ref{eq1}), we have

\begin{equation}
\label{eq3} g = \det (g_{\mu \nu } ) = - \frac{1}{\Xi ^2}\rho ^4\sin
^2\theta .
\end{equation}
The contravariant variant of $g_{\mu \nu } $ is

\[
\frac{\partial ^2}{\partial s^2} = - \frac{1}{\rho ^2\Delta _r
\Delta _\theta }\left[ {\Delta _\theta (r^2 + a^2)^2 - \Delta _r
a^2\sin ^2\theta } \right]
\]

\[\frac{\partial ^2}{\partial t^2} +
\frac{\Delta _r }{\rho ^2}\frac{\partial ^2}{\partial r^2} +
\frac{\Delta _\theta }{\rho ^2}\frac{\partial ^2}{\partial \theta
^2}+ \frac{\Xi ^2}{\Delta _r \Delta _\theta \rho ^2\sin ^2\theta
}\left( {\Delta _r - \Delta _\theta a^2\sin ^2\theta } \right)
\]

\begin{equation}
\label{eq4}
 \frac{\partial ^2}{\partial \varphi ^2} - \frac{2\Xi a}{\rho
^2\Delta _r \Delta _\theta }\left[ {\Delta _\theta (r^2 + a^2) -
\Delta _r } \right]\frac{\partial ^2}{\partial t\partial \varphi }.
\end{equation}
The electromagnetic potential is [39]

\begin{equation}
\label{eq5} A_\mu = - \frac{qr}{\rho ^2}\left( {1,0,0, - \frac{a\sin
^2\theta }{\Xi }} \right).
\end{equation}
The horizon surface equation of the Kerr-Newman-de Sitter spacetime
is

\begin{equation}
\label{eq6} \Delta _r = - \frac{1}{3}\Lambda (r - r_c )(r - r_ + )(r
- r_ - )(r - r_{ - - } ) = 0.
\end{equation}
When $\frac{1}{\Lambda } > > M^2 > a^2 + Q^2$, equation $\Delta _r =
0$ has $r_c $, $r_ + $, $r_ - $ and $r_{ - \, - }$ four real
solutions, where $r_c $, $r_ + $, $r_ - $ are positive and
$r_c>r_+$, $r_{ - \, - } $ is negative. $r_c $, $r_{ - \, - }
$correspond to de Sitter cosmological horizon. $r_ + $, $r_ - $
correspond to Kerr-Newman black hole horizon.

The Abbott and Deser (AD) mass of the KNdS solution can be expressed
in terms of the horizon radius $r_ + $, $a$ and charge $q$ [37-39]:

\begin{equation}
\label{eq7} E = \frac{M}{\Xi } = \frac{(r_ + ^2 + a^2)(r_ + ^2 - 3 /
\Lambda ) - q^23 / \Lambda }{2\Xi r_ + 3 / \Lambda }.
\end{equation}
The Hawking temperature of the black hole horizon is given by

\begin{equation}
\label{eq8} T_ + = \frac{1}{4\pi }\frac{\Delta _r '(r_ + )}{r_ + ^2
+ a^2}
 = - \frac{3r_ + ^4 + r_ + ^2 (a^2 - 3 / \Lambda ) + (a^2 + q^2)3 / \Lambda
}{4\pi r_ + (r_ + ^2 + a^2)3 / \Lambda }.
\end{equation}
The entropy associated with the black hole horizon can be calculated
as

\begin{equation}
\label{eq9} S_ + = \frac{\pi (r_ + ^2 + a^2)}{\Xi }.
\end{equation}
The angular velocity of the black hole horizon is given by

\begin{equation}
\label{eq10} \Omega _ + = \frac{a\Xi }{r_ + ^2 + a^2}.
\end{equation}
The angular momentum $J$, the electric charge $Q$, and the electric
potential $\phi _ + $ are given by

\begin{equation}
\label{eq11} J = \frac{Ma}{\Xi ^2}, \quad Q = \frac{q}{\Xi }, \quad
\phi _ + = \frac{qr_ + }{r_ + ^2 + a^2}.
\end{equation}
The quantities obtained above of the black hole horizon satisfy the
first law of thermodynamics:

\begin{equation}
\label{eq12} dE = T_ + dS_ + + \Omega _ + dJ + \phi _ + dQ.
\end{equation}
The cosmological horizon has associated thermodynamic quantities

\[
T_c = \frac{ - 1}{4\pi }\frac{\Delta _r '(r_c )}{r_c^2 + a^2}
 = \frac{3r_c^4 + r_c^2 (a^2 - 3 / \Lambda ) + (a^2 + q^2)3 / \Lambda }{4\pi
r_c (r_c^2 + a^2)3 / \Lambda },
\]

\[
S_c = \frac{\pi (r_c^2 + a^2)}{\Xi }, \quad \Omega _c =
\frac{ - a\Xi }{r_c^2 + a^2},
\]

\begin{equation}
\label{eq13}  \quad J = \frac{Ma}{\Xi ^2}, \quad Q = \frac{q}{\Xi },
\quad \phi _c = \frac{ - qr_c }{r_c^2 + a^2}.
\end{equation}
The Balasubranmanian, de Boer and Minic (BBM) mass of KNdS is

\begin{equation}
\label{eq14} \tilde {E} = \frac{ - M}{\Xi } = \frac{(r_c^2 +
a^2)(r_c^2 - 3 / \Lambda ) - q^23 / \Lambda }{2\Xi r_c 3 / \Lambda
}.
\end{equation}
The quantities obtained above of the cosmological horizon also
satisfy the first law of thermodynamics:

\begin{equation}
\label{eq15} d\tilde {E} = T_c dS_c + \Omega _c dJ + \phi _c dQ.
\end{equation}
\section{Klein-Gordon equation}
In curved spacetime, the Klein-Gordon equation of a charged
particles is

\begin{equation}
\label{eq16} \frac{1}{\sqrt { - g} }\left[ {\left( {\frac{\partial
}{\partial x^\mu } - ieA_\mu } \right)\sqrt { - g} g^{\mu \nu
}\left( {\frac{\partial }{\partial x^\nu } - ieA_\nu } \right)\Phi }
\right] - \mu _0^2 \Phi = 0,
\end{equation}

\noindent where $\mu _0 $ is the mass of the scalar particle, $e$ is
the charge of the particle. Substituting (\ref{eq4}) and (\ref{eq5})
into (\ref{eq16}), we have

\[
\frac{1}{\Delta _r \Delta _\theta }\left[ {\Delta _\theta (r^2 +
a^2)^2 - \Delta _r a^2\sin ^2\theta } \right]\frac{\partial ^2\Phi
}{\partial t^2}
\]

\[+ 2i\frac{eqr}{\Delta _r }\left[ {a\Xi \frac{\partial }{\partial \varphi }
+ (r^2 + a^2)\frac{\partial }{\partial t}} \right]\Phi
\]

\[
- \frac{\Xi ^2}{\Delta _r \Delta _\theta \sin ^2\theta }\left( {\Delta _r -
\Delta _\theta a^2\sin ^2\theta } \right)\frac{\partial ^2\Phi
}{\partial \varphi ^2}
\]

\[
 + \frac{2\Xi a}{\Delta _r \Delta _\theta
}\left[ {\Delta _\theta (r^2 + a^2) - \Delta _r }
\right]\frac{\partial ^2\Phi }{\partial t\partial \varphi }-
\frac{\partial }{\partial r}\Delta _r \frac{\partial \Phi }{\partial
r}
\]

\begin{equation}
\label{eq17}
- \frac{1}{\sin \theta }\frac{\partial }{\partial
\theta }\left( {\sin \theta \Delta _\theta } \right)\frac{\partial
\Phi }{\partial \theta }
 = \left[ { - \mu _0^2 \rho ^2 + e^2q^2r^2\frac{1}{\Delta _r }} \right]\Phi
.
\end{equation}
Separating variable $t$ and $(r,\theta )$, and let

\begin{equation}
\label{eq18} \Phi (t,r,\theta ,\varphi ) = e^{ - i\omega
t}e^{im\varphi }\psi (r,\theta ),
\end{equation}
(\ref{eq17}) can be reduced to

\[
 - \omega ^2\frac{1}{\Delta _r \Delta _\theta }\left[ {\Delta _\theta (r^2 +
a^2)^2 - \Delta _r a^2\sin ^2\theta } \right]\psi
\]

\[
+ 2\frac{eqr}{\Delta _r }\left[ {(r^2 + a^2)\omega - a\Xi m} \right]\psi
\]

\[
+ m^2\frac{\Xi ^2}{\Delta _r \Delta _\theta \sin ^2\theta }\left( {\Delta
_r - \Delta _\theta a^2\sin ^2\theta } \right)\psi
\]

\[
 + m\omega\frac{2\Xi a}{\Delta _r \Delta _\theta }\left[ {\Delta _\theta (r^2
+ a^2) - \Delta _r } \right]\psi- \frac{\partial }{\partial r}\Delta
_r \frac{\partial \psi }{\partial r}
\]

\begin{equation}
\label{eq19}
- \frac{1}{\sin \theta }\frac{\partial }{\partial
\theta }\left( {\sin \theta \Delta _\theta } \right)\frac{\partial
\psi }{\partial \theta }
 = \left[ { - \mu _0^2 \Sigma + e^2q^2r^2\frac{1}{\Delta _r }} \right]\psi
.
\end{equation}
Let $\psi = \chi (\theta )R(r)$, we have

\[
\frac{1}{\sin \theta }\frac{\partial }{\partial \theta
}(\sin \theta \Delta _\theta )\frac{\partial }{\partial \theta }\chi
(\theta )
\]

\begin{equation}
\label{eq20}  = \left[ {\frac{1}{\Delta _\theta }\left( {\omega
a\sin \theta - \frac{m\Xi }{\sin \theta }} \right)^2 + \mu _0
a^2\cos ^2\theta - \lambda } \right]\chi (\theta ),
\end{equation}

\[
 \frac{d}{dr}\Delta _r \frac{d}{dr}R(r)
\]

\begin{equation}
\label{eq21}= \left[ {\lambda + \mu _0^2 r^2 - \frac{1}{\Delta _r
}[\omega (r^2 + a^2) - am\Xi - eqr]^2} \right]R(r).
\end{equation}

\noindent where $\lambda $ is separation variable constant, $\omega
$ is the energy of the radiation particles, $m$ is the projection of
the angular momentum of the radiation particle on the rotation axis.
$e$ is the charge of the radiation particle. Letting $K = (r^2 +
a^2)\omega - am\Xi - eqr$, Eq.(\ref{eq21}) can be reduced to

\[
\Delta _r \frac{d^2R(r)}{dr^2} + 2\left( {r - \frac{2}{3}\Lambda r^3
- \frac{1}{3}\Lambda ra^2 - M} \right)\frac{dR(r)}{dr}
\]

\begin{equation}
\label{eq22} = \left[ {\lambda + \mu _0^2 r^2 - \frac{K^2}{\Delta _r
}} \right]R(r).
\end{equation}
\section{Tortoise coordinate transformation}
Now we introduce the tortoise coordinate $r_\ast$ through the
following equations

\[
dr_\ast = \frac{1}{\Delta _r }(r^2 + a^2)dr,
\]

\[
\frac{d}{dr} = \frac{r^2 + a^2}{\Delta _r }\frac{d}{dr_\ast },
\]

\[
\frac{d^2}{dr^2} = \left( {\frac{r^2 + a^2}{\Delta _r }}
\right)^2\frac{d^2}{dr_\ast ^2 }
\]

\begin{equation}
\label{eq23}  +2\frac{Q^2r - Mr^2 + Ma^2 + (r^2 + a^2)^2\Lambda r /
3}{\Delta _r^2 }\frac{d}{dr_\ast }.
\end{equation}
Substituting (\ref{eq23}) into (\ref{eq22}), we have

\begin{equation}
\label{eq24} (r^2 + a^2)^2\frac{d^2R(r)}{dr_\ast ^2 } + 2r\Delta _r
\frac{dR(r)}{dr_\ast } = \left[ {\Delta _r (\lambda + \mu _0^2 r^2)
- K^2} \right]R(r).
\end{equation}
Near the black hole horizon $\Delta _r (r_ + ) \to 0$, so
(\ref{eq24}) can be reduced to

\begin{equation}
\label{eq25} \frac{d^2R(r)}{dr_\ast ^2 } + (\omega - \omega _0
)^2R(r) = 0,
\end{equation}

\noindent where $\omega _0 = m\Omega _ + + e\phi _ + $. The solution
of (\ref{eq25}) is

\begin{equation}
\label{eq26} R = e^{\pm i(\omega - \omega _0 )r_\ast }.
\end{equation}
Considering time factor, near the black hole horizon $r_ + $ this
solution is

\begin{equation}
\label{eq27} \Psi _{out} = e^{ - i\omega t\pm i(\omega - \omega _0
)r_\ast }.
\end{equation}
Letting $\hat {r} = \frac{\omega - \omega _0 }{\omega }r_\ast $, on
the black hole horizon surface we derive the ingoing wave solution

\begin{equation}
\label{eq28} \Psi _{in} = e^{ - i\omega (t + \hat {r})} = e^{ -
i\omega v},
\end{equation}

\noindent and outgoing wave solution

\begin{equation}
\label{eq29} \Psi _{out} (r > r_ + ) = e^{ - i\omega (t - \hat {r})}
= e^{ - i\omega v}e^{2i\omega \hat {r}}
 = e^{ - i\omega v}e^{2i(\omega - \omega _0 )r_\ast },
\end{equation}

\noindent where $v = t + \hat {r}$ is Eddington-Finkelstein
coordinate. Because of $\frac{dr}{\Delta _r } = \frac{dr_\ast }{r^2
+ a^2}$, near the black hole horizon surface $r_ + $, we have

\begin{equation}
\label{eq30} \ln (r - r_ + ) = \frac{1}{r_ + ^2 + a^2}\left.
{\frac{d\Delta _r }{dr}} \right|_{r = r_ + } r_\ast
 = 2\kappa _+ r_\ast ,
\end{equation}

\noindent where

\begin{equation}
\label{eq31} \kappa _ + = - \frac{3r_ + ^4 + r_ + ^2 (a^2 - 3 /
\Lambda ) + (a^2 + q^2)3 / \Lambda }{r_ + (r_ + ^2 + a^2)3 / \Lambda
},
\end{equation}
$\kappa _ + $ is gravitational acceleration on the black hole
horizon surface $r_ + $. From (\ref{eq30}), we have

\begin{equation}
\label{eq32} (r - r_ + ) = \exp (2\kappa _ + r_\ast ),
\end{equation}

\noindent and the outgoing wave is rewritten as

\begin{equation}
\label{eq33} \Psi _{out} (r > r_ + ) = e^{i\omega v}(r - r_ +
)^{\textstyle{i \over {\kappa _ + }}(\omega - \omega _0 )}.
\end{equation}
Obviously, Eq.(\ref{eq33}) has singularity on horizon surface $r_ +
$, and can only describe outgoing particles outside horizon $r_ + $.
It can not describe the outgoing particles on the horizon.
\section{Analytic extension}
We are interested in outing wave when the black hole radiation is
studied. From (\ref{eq33}), the outing wave is singular at $r = r_ +
$, we can extend $\Psi _{out} $ from the outside of the black hole
into the inside of the black hole. We take the singularity $r = r_ +
$ as the center of a circle, and take $\left| {r - r_ + } \right|$
as radius. By analytical continuation rotating $ - \pi $ through the
lower-half complex $r$ plane, we have

\begin{equation}
\label{eq34} (r - r_ + ) \to \left| {r - r_ + } \right|e^{ - i\pi }
= (r_ + - r)e^{ - i\pi }.
\end{equation}
So we obtain the outgoing wave in the horizon surface $r_ + $,

\[
\Psi _{out} (r < r_ + ) = e^{i\omega v}(r_ + - r)^{\textstyle{i
\over {\kappa _ + }}(\omega - \omega _0 )}e^{\textstyle{\pi \over
{\kappa _ + }}(\omega - \omega _0 )}
\]

\begin{equation}
\label{eq35}
= e^{\pi (\omega - \omega _0 ) / \kappa _ + }e^{ -
i\omega v}e^{2i(\omega - \omega _0 )r_\ast }.
\end{equation}
Eqs.(\ref{eq35}) and (\ref{eq29}) describe the outgoing wave of
outside and inside of black hole, respectively. So, for outgoing
wave of particle with energy $\omega $, charged $e$ and angular
momentum $m$, the outgoing rate at the horizon surface is given by

\begin{equation}
\label{eq36} \Gamma _ + = \left| {\frac{\Psi _{out} (r > r_ +
)}{\Psi _{out} (r < r_ + )}} \right|^2 = e^{ - 2\pi (\omega - \omega
_0 ) / \kappa _+ }.
\end{equation}
Near the the cosmological horizon $\Delta _c (r_c) \to 0$, by the
same method, we can solve equation (\ref{eq25}) and derive the
outgoing wave of particle with energy $\omega$, charged $e$ and
angular momentum $m$, the outgoing rate at the cosmological horizon
surface is given by

\begin{equation}
\label{eq37} \Gamma _c = \left| {\frac{\Psi _{out} (r < r_c )}{\Psi
_{out} (r > r_c )}} \right|^2 = e^{2\pi (\omega + \omega _c ) /
\kappa _c },
\end{equation}

\noindent where $\omega _c = m\Omega _c + e\phi _c $,

\begin{equation}
\label{eq38} \kappa _c = \frac{3r_c^4 + r_c^2 (a^2 - 3 / \Lambda ) +
(a^2 + q^2)3 / \Lambda }{r_c (r_c^2 + a^2)3 / \Lambda },
\end{equation}
$\kappa _c$ is gravitational acceleration on the cosmological
horizon surface.
\section{Radiation spectrum}

According to the above discussion, we obtain that the total energy,
angular momentum and charge of the spacetime are respectively $M /
\Xi + \omega $, $J + m$ and $Q + e$ [40]. However, between the black
hole event horizon and the cosmological horizon the energy, angular
momentum and charge of the radiation particles are respectively
$\omega $, $m$ and $e$. Before radiation the energy of the black
hole is $M / \Xi + \omega $, the charge is $Q + e$, and angular
momentum is $J + m$. So we can take the process of the black hole
that radiates particles as the process that the Kerr-Newman-de
Sitter black hole transfers from the initial state (energy $M / \Xi
+ \omega $, charge $Q + e$ and angular momentum $J + m)$ to the
final state (energy $E$, charge $Q$ and angular momentum $J)$.

Now we replace the energy, angular momentum and charge of the
radiation particles with the parameters of the Kerr-Newman-de Sitter
black hole. This result embodies the reaction of the radiation to
spacetime. When parameters of the Kerr-Newman-de Sitter black hole
are used, we must guarantee that the total energy, angular momentum
and charge of spacetime are all conserved. That is

\[
 - \omega = \Delta E,
\quad \omega = \Delta \tilde {E}
\]

\begin{equation}
\label{eq39}
 - e = \Delta Q,
\quad
 - m = \Delta J,
\end{equation}

\noindent where $\Delta E$ or $\Delta \tilde {E}$, $\Delta Q$ and
$\Delta J$ is the change of energy, charge and angular momentum of
the black hole event horizon and cosmological horizon before and
after radiation, respectively. Substituting (\ref{eq39}) and
(\ref{eq12}) into (\ref{eq36}), we obtain the outgoing rate of the
outgoing wave on the black hole horizon

\begin{equation}
\label{eq40} \Gamma _ + = e^{\Delta S_ + }.
\end{equation}
Substituting (\ref{eq39}) and (\ref{eq15}) into (\ref{eq37}), we
derive the outgoing rate of the outgoing wave on the cosmological
horizon surface

\begin{equation}
\label{eq41} \Gamma _c = e^{\Delta S_c }.
\end{equation}
Taking the black hole as a thermodynamic system, we discuss the
radiation rate that this thermodynamic system radiates particles
with energy $\omega $, angular momentum $m$ and charge $e$.
According to the discussion, when we do not consider the radiation
of cosmological horizon, the emission rate that the black hole event
horizon radiates particles with energy $\omega $, angular momentum
$m$ and charge $e$ is (\ref{eq40}). After the black hole event
horizon radiates particles with energy $\omega $, angular momentum
$m$ and charge $e$, from (\ref{eq15}), the change of the entropy
corresponding to the cosmological horizon is $\Delta S_c $. So we
can take (\ref{eq40}) as the probability of the change of the
entropy corresponding to the cosmological horizon caused by
radiating particles from the black hole horizon. By the same method
we can take (\ref{eq41}) as the probability of the change of the
entropy corresponding to the black hole horizon caused by radiating
particles from the cosmological horizon. Thus for the black hole
event horizon because of radiating particles with energy $\omega $,
angular momentum $m$ and charge $e$ there are two way that cause the
change of the entropy $\Delta S_ + $. One way is the black hole
radiates particles with energy $\omega $, angular momentum $m$ and
charge $e$. The probability is given by (\ref{eq40}). The other is
the cosmological horizon radiates particles with energy $\omega $,
angular momentum $m$ and charge $e$. And the probability is given by
(\ref{eq41}). So for the black hole event horizon because of
radiating particles with energy $\omega $, angular momentum $m$ and
charge $e$, the probability that entropy change is $\Delta S_ + $

\begin{equation}
\label{eq42} \Gamma = \Gamma _ + \Gamma _c = e^{\Delta S_ + + \Delta
S_c }.
\end{equation}
It is known that the radiation spectrum of Kerr-Newman-de Sitter
black hole is related not only to the change of the entropy of the
black hole horizon but also to the one of the cosmological horizon.
And the radiation spectrum satisfies the unitary principle.
\section{Conclusion and Discussion}

For de Sitter, there is radiation not only from the black hole event
horizon but also from the cosmological horizon. Refs.[31,41-43] took
the black hole event horizon and the cosmological horizon as two
independent horizons, when they studied the quantum tunneling of
those spacetimes. They discussed the radiation spectrums
respectively and did not consider the relation between the black
hole event horizon and the cosmological horizon.

We extend the classical Damour-Ruffini method and discuss Hawking
radiation spectrum in the Kerr-Newman-de Sitter(KNdS) black hole
under the condition that the total energy, angular momentum and
charge of spacetime are conserved. We discuss the particle radiation
with arbitrary angular momentum. After considering the reaction to
spacetime from radiation particles, the radiation spectrum of the
Kerr-Newman-de Sitter black hole is related not only to the entropy
change of the black hole horizon but also to the entropy change of
the cosmological horizon. The black hole radiation spectrum
satisfies the unitary principle. Since

\begin{equation}
\label{eq43} \Delta J = \Delta (Ma / \Xi ^2) = - m,
\end{equation}
where $m$ takes only zero or integer values, the black hole angular
momentum $J$ is quantized. From (\ref{eq43}), the change of $m$ is
determined not only by the change of $M$ but also the change of the
black hole rotating parameter $a$. The research on axisymmetric
black hole in Refs.[44,45] is under the condition that $a$ is
constant.

When only consider the tunneling radiation of the black hole event
horizon, we adopt the same method to Ref.[34]. Starting from
Damour-Ruffini method, we discuss the outgoing wave of particles
with energy $\omega $, charge $e$ and angular momentum $m$. Our
result is (\ref{eq40}) ((\ref{eq19}) in Ref.[34]). In Ref.[34] the
radiation process was taken as a integration process. Summing up the
energy of radiation particles, they derived that the radiation
spectrum departed from the black body spectrum. Under the condition
that the total energy, angular momentum and charge of spacetime are
conserved, we obtain the probability that the black hole transfers
from initial state (energy $M / \Xi + \omega $, charge $Q + e$ and
angular momentum $J + m)$ to final state (energy $M / \Xi $, charge
$Q$ and angular momentum $J)$. That is the probability the black
hole event horizon radiates particles with energy $\omega $, charge
$e$ and angular momentum $m$. Our result is consistent with that
obtained by Parikh and Wilczek.

Because the Kerr-Newman-de Sitter black hole has the event horizon
and the cosmological horizon, and the state parameters that describe
two horizons are the same, so the radiations of two horizons are
correlative. When we discuss the radiation spectrums of those black
holes, we must consider the relevance between the black hole event
horizon and the cosmological horizon. In this paper, we consider the
radiation spectrums of two correlative
horizons.\vspace{0.25cm}\\
\textbf{Acknowledgment}

This project was supported by the Shanxi Natural Science Foundation
of China under Grant No. 2006011012 and the Doctoral Scientific
Research Starting Foundation of Shanxi Datong University, China.

\end{document}